\documentclass[11pt]{article}

\date{}

\usepackage[utf8]{inputenc} 
\usepackage[T1]{fontenc}    
\usepackage{hyperref}       
\usepackage{url}            
\usepackage{booktabs}       
\usepackage{amsfonts}       
\usepackage{nicefrac}       
\usepackage{microtype}      
\usepackage{xspace}
\usepackage{amsmath}
\usepackage{authblk}
\usepackage{siunitx}
\usepackage{comment}
\usepackage{amsmath}
\usepackage{amsthm}
\usepackage{graphicx}
\usepackage{rotating}
\usepackage{subcaption} 
\usepackage{amssymb}
\usepackage{autobreak}
\usepackage{mathtools}
\usepackage{xcolor}
\usepackage{bbm}
\usepackage{algorithm}
\usepackage{algorithmic}

\newtheorem{prop}{Proposition}

\makeatletter

\usepackage[numbers,sort&compress]{natbib}

\title{SPIDER: Scalable Probabilistic Inference for Differential Earthquake Relocation}

\author[1]{Zachary E. Ross}
\author[1]{John D. Wilding}
\author[2]{Kamyar Azizzadenesheli}
\author[3]{Aitaro Kato}

\affil[1]{Seismological Laboratory, California Institute of Technology}
\affil[2]{Nvidia Corporation, Santa Clara, California}
\affil[3]{Earthquake Research Institute, University of Tokyo}

\begin{document}
\maketitle

\begin{abstract}
Seismicity catalogs are larger than ever due to an explosion of techniques for enhanced earthquake detection and an abundance of high-quality datasets. Bayesian inference is an appealing framework for locating earthquakes due to its ability to propagate and quantify uncertainty into the inversion results, but traditional methods do not scale well to high-dimensional parameter spaces, making them unsuitable for double-difference relocation where the number of parameters can reach the millions. Here we introduce SPIDER, a scalable Bayesian inference framework for double-difference hypocenter relocation. SPIDER uses a physics-informed neural network Eikonal solver together with a highly efficient sampler called Stochastic Gradient Langevin Dynamics to generate posterior samples jointly for entire seismicity catalogs. We show that traditional double-difference relocation formulations neglect residual correlation between observations with common events, which biases uncertainty estimates. Our formulation is designed to whiten this residual correlation, and is readily parallelized over multiple GPUs for enhanced computational efficiency. We demonstrate the capabilities of SPIDER on a rigorous synthetic seismicity catalog and three real data catalogs from California and Japan. We introduce several ways to analyze high-dimensional posterior distributions to aid in scientific interpretation and evaluation.
\end{abstract}

\section*{Plain Language Summary}
We develop a technique called SPIDER for locating large volumes of earthquakes precisely while focusing on quantifying the uncertainty in the locations. We benchmark the method on four datasets of earthquakes collected from California and Japan against known baseline methods. The method is designed to work efficiently for very large datasets. We also introduce several ways in which the earthquake locations, together with the uncertainties, can be visualized to better interpret features such as faults below the surface.

%
%

\section{Introduction}
Earthquake hypocenters are an essential component of numerous scientific and hazard analyses, including fault zone characterization \citep{plesch_community_2007}, studies of earthquake sequences \citep{munchmeyer_seismic_2025,trugman_coherent_2023}, travel time tomography \citep{bennington_three-dimensional_2008}, and much more. Access to precise hypocenters and reliable estimates of their uncertainty has been an essential part of advances in earthquake science over the last several decades \citep{lomax_probabilistic_2000}. Traditional hypocenter estimates are individually determined for each event and are referred to as absolute locations. These locations are generally limited by the accuracy of a velocity model and the timing accuracy of phase picks. To overcome both limitations, double-difference methods were developed \citep{waldhauser_double-difference_2000,gillard_highly_1996,lin_tests_2005,hauksson_waveform_2012,trugman_growclust:_2017}, which rely on information between pairs of events to jointly determine hypocenters for an entire catalog. Double-difference methods overcome limitations in knowledge about velocity structure by effectively canceling out unknown structure that happens to be common to multiple events \citep{waldhauser_double-difference_2000}. They further overcome limitations in picking accuracy by allowing for more precise timing measurements between pairs of event waveforms from cross-correlation \citep{rubin_streaks_1999}.  Double-difference methods have been applied extensively to great success and represent the state-of-the-art in location quality in many scenarios.

In addition to location accuracy, the notion of uncertainty has become an important part of the conversation to allow proper scientific interpretability \citep{tarantola_inverse_1981,lomax_probabilistic_2000}. This has been formulated most rigorously with Bayesian inference to allow for complete probabilistic descriptions of the uncertainties and correlations associated with the hypocentral parameters \citep{lomax_high-precision_2022}. Bayesian inference makes the assumptions in the probability model explicit, allowing for a clear interpretation of the solutions \citep{zhang_3-d_2023,duputel_accounting_2014,smith_hyposvi_2021,ragon_accounting_2018}. The primary downside is the need to compute not just one single solution, but rather an ensemble (or distribution) of solutions, which is in most cases a harder problem and more computationally demanding.

As waveform datasets have grown in size and detection methods have improved significantly \citep[e.g.]{zhu_phasenet:_2019,sun_phase_2023}, seismicity catalogs are routinely at the size of $10^5-10^6$ events today with even more massive volumes of observations \citep{ross_searching_2019,sippl_northern_2023,poggiali_fault_2025}. Such catalogs pose challenges for the more established double-difference methods in terms of scalability \citep{waldhauser_double-difference_2000,trugman_growclust:_2017}. More recently, a method using graph neural networks was developed to overcome limitations in dataset size \citep{mcbrearty_double_2025}. This approach is a significant step forward in this regard, but is still focused on obtaining deterministic solutions through minimizing an L1 norm of residuals.

In this study, our contributions are as follows. We introduce Scalable Probabilistic Inference for Differential Earthquake Relocation (SPIDER), an algorithm designed to solve for the full probabilistic solution for a double-difference catalog at scale. SPIDER uses a highly efficient algorithm called Stochastic Gradient Langevin Dynamics to generate high-dimensional full posterior samples using a physics-informed neural network Eikonal solver. This allows for uncertainties and correlations to be rigorously characterized both between and within events. We introduce several novel ways in which the complete probabilistic solution for such high-dimensional datasets can be visualized for scientific interpretation. We demonstrate the capabilities of SPIDER on three real data catalogs from California and Japan, and one synthetic benchmark dataset.
\section{Background}
\subsection{Preliminaries}

\noindent An earthquake hypocenter, $x = (x^s, y^s, z^s) \in \mathbb{R}^3$,  is the location at which an earthquake nucleates. Let us define $s = (x^r,y^r,z^r) \in \mathbb{R}^3$ as the location at which an observation is made. The travel time of a seismic wave from $x$ to $s$ is given by $T_\alpha(x; s) = g_\alpha(x, s)$, where $g_\alpha$ is a nonlinear operator describing a physical forward model for seismic P-waves or S-waves. Although we rarely observe $T_\alpha$ directly, we can measure the absolute arrival time $t^\alpha(s) = g_\alpha(x, s)+ t^s$, where $t^s$ is the origin time of the earthquake. Traditional hypocenter methods attempt to infer $(x, t^s)$ given a set of observed arrival times, $\{t(s_i)\}_{i=1}^N $.  In this inverse problem, errors in fitting the observations can result from either measurement (picking) errors or uncertainty in the forward model due to imperfect knowledge about the velocity structure. 

The double-difference method is a pairwise earthquake relocation strategy that has been widely successful \citep{waldhauser_double-difference_2000,waldhauser_large-scale_2008,matoza_comprehensive_2021,hauksson_waveform_2012}. Fundamentally, it provides superior resolution over absolute location methods for two reasons. First, differential arrival times can generally be measured with cross-correlation to a greater degree of precision than absolute arrival times measured by humans or machine learning algorithms. Second, the forward model for the double-difference relocation problem is, under certain conditions, effectively independent of the velocity model used, except for in the near-source region.

\noindent Let $i,j$ denote the indices of two events and let $T(x) \equiv T_\alpha(x; s)$ be a scalar field on $\mathbb{R}^3$ for some fixed $s$. Then the differential arrival time between these events at a station $s$ is
\begin{equation}
   \tau_{ij} (s) \equiv t_i^\alpha(s) - t_j^\alpha(s)
   = T_\alpha(x_i; s) - T_\alpha(x_j; s)  + t_i^s - t_j^s.
\end{equation}

\begin{prop}
 Let $\mathbf{H}_T(x)$ denote the Hessian of $T$. Then
 \begin{equation}
     T(x_j) - T(x_i) \approx \nabla_{x} T(x_i)\cdot (x_j - x_i)
     \quad \text{if} \quad
     \|x_j - x_i\| \ll \frac{\|\nabla_{x} T(x_i)\|}{\|\mathbf{H}_T(x_i)\|}.
     \label{eq:dd_linearization}
 \end{equation}
 In particular, if $T$ satisfies the eikonal equation $\|\nabla_{x} T(x)\| = 1/c(x)$, this becomes
 \[
 \|x_j - x_i\| \ll \frac{1}{c(x_i)\,\|\mathbf{H}_T(x_i)\|}.
 \]
\end{prop}
\begin{proof}
If $T$ is a smooth function of $x$, we can write its Taylor expansion about some point $x_0$ as 
\begin{equation}
\begin{aligned}
T(x_i) &= 
T(x_0)
+ \nabla_{x} T(x_0)\cdot(x_i-x_0)
+ \tfrac{1}{2}(x_i-x_0)^\top \mathbf{H}_T(x_0)(x_i-x_0)
+ \mathcal{O}(\|x_i-x_0\|^3), \\
T(x_j) &= 
T(x_0)
+ \nabla_{x} T(x_0)\cdot(x_j-x_0)
+ \tfrac{1}{2}(x_j-x_0)^\top \mathbf{H}_T(x_0)(x_j-x_0)
+ \mathcal{O}(\|x_j-x_0\|^3),
\end{aligned}
\label{eq:full-taylor}
\end{equation}
where $\mathbf{H}_T(x_0)$ is the Hessian of $T$ at $x_0$. Taking the difference of these two equations gives
\begin{equation}
\begin{aligned}
T(x_j) - T(x_i)
&=
\nabla_{x} T(x_0)\cdot(x_j - x_i) \\
&\quad
+ \tfrac{1}{2}\big[
(x_j-x_0)^\top \mathbf{H}_T(x_0)(x_j-x_0)
- 
(x_i-x_0)^\top \mathbf{H}_T(x_0)(x_i-x_0)
\big] \\
&\quad
+ \mathcal{O}(\|x_j-x_0\|^3) - \mathcal{O}(\|x_i-x_0\|^3).
\end{aligned}
\label{eq:full-diff-reversed}
\end{equation}
Setting $x_0 = x_i$ yields the second-order Taylor expansion about $x_i$:
\begin{equation}
\begin{aligned}
T(x_j) - T(x_i)
&=
\nabla_{x} T(x_i)\cdot(x_j - x_i)
+ \tfrac{1}{2}
(x_j-x_i)^\top \mathbf{H}_T(x_i)(x_j-x_i)
+ \mathcal{O}(\|x_j-x_i\|^3).
\end{aligned}
\label{eq:full-final-diff}
\end{equation}
Let $\Delta x = x_j - x_i$. Then
\[
T(x_j) - T(x_i)
=
\nabla_{x} T(x_i)\cdot\Delta x
+ \tfrac{1}{2}
\Delta x^\top \mathbf{H}_T(x_i)\,\Delta x
+ \mathcal{O}(\|\Delta x\|^3).
\]
By the Cauchy--Schwarz inequality, the magnitude of the first-order term is bounded by
\[
\| \nabla_{x} T(x_i)\cdot\Delta x\|
\le \|\nabla_{x} T(x_i)\|\,\|\Delta x\|.
\]
Using the quadratic–form inequality,
\[
\left\|
\tfrac{1}{2}
\Delta x^\top \mathbf{H}_T(x_i)\,\Delta x
\right\|
\le
\tfrac{1}{2}\|\mathbf{H}_T(x_i)\|\,\|\Delta x\|^2.
\]
Thus a sufficient condition for the second-order and higher terms to be negligible compared to the first-order term is
\begin{equation}
\tfrac{1}{2}\|\mathbf{H}_T(x_i)\|\,\|\Delta x\|^2 
\ll 
\|\nabla_{x} T(x_i)\|\,\|\Delta x\|
\;\;\Longrightarrow\;\;
\|\Delta x\|
\ll
\frac{2\,\|\nabla_{x} T(x_i)\|}{\|\mathbf{H}_T(x_i)\|}.
\label{eq:dd_criterion}
\end{equation}
Ignoring order-one constants, this yields \eqref{eq:dd_linearization}.
\end{proof}

For a homogeneous medium with constant speed $c$, the travel time is
\[
T(x;s) = \frac{1}{c}\|x - s\| = \frac{r}{c},
\]
where $r = \|x-s\|$ and $r = x - s$. The gradient is
\[
\nabla_{x} T = \frac{r}{cr}.
\]
The Hessian is
\[
\mathbf{H}_T = \frac{1}{cr}\left(\mathbf{I} - \hat{r}\hat{r}^\top\right),
\]
where $\hat{r} = r/r$ and $\mathbf{I}$ is the identity matrix. The matrix $(\mathbf{I} - \hat{r}\hat{r}^\top)$ is the orthogonal projector onto the subspace perpendicular to $\hat{r}$ and therefore has operator norm~$1$. Hence
\[
\|\mathbf{H}_T\| = \frac{1}{cr}.
\]
In this homogeneous case, the double-difference criterion \eqref{eq:dd_criterion} reduces to
\[
\|\Delta x\| \ll \|x - s\|,
\]
i.e., the event separation must be small compared to the source–receiver distance. It is worth noting that in the linearization regime, the error in the forward model can be written as
\begin{equation}
    b_n \approx \delta s(x_i) \cdot (x_j - x_i),
    \label{eq:fwd_model_error}
\end{equation}
where $\delta s$ is the slowness error vector at $x_i$. As such, the error in the forward model increases linearly with event separation, which we will later show introduces separation-dependent correlation in the residuals. At the same time, in this regime the error has a particularly simple form, which is much easier to model without resorting to full tomographic methods.

While the linearized forward model could be used to solve the inverse problem, its assumptions may not always be satisfied in practice and can lead to artifacts. Therefore we focus on restricting the set of observations to those satisfying eq. \ref{eq:dd_linearization}, to simplify and limit the forward model error, but still employ the full non-linear forward model, i.e.,
\begin{equation}
    \tau_n(\theta) = t_i^{\alpha_n}(s_n) - t_j^{\alpha_n}(s_n) = g_{\alpha_n}(x_i,x_j,s_n, t_i^s, t_j^s) .
    \label{eq:forwardmodel}
\end{equation}
Given observations $\mathbf{\tau}^{obs} \in \mathbb{R}^N$,  and defining $\theta = \{(x_m, t_m^s)\}_{m=1}^M$, we can formulate a basic inverse problem as,
\begin{equation}
\operatorname*{arg\,min}_{\theta} \, L\left(\mathbf{\tau}^{obs}, \tau(\theta)\right)
\label{eq:stage1loss}
\end{equation}
where $L$ is some objective function.

\subsection{Related Work}
\textbf{HypoDD}. The HypoDD method \citep{waldhauser_double-difference_2000} is an iterative linearized inversion scheme for relative event relocation in which the forward model is explicitly linearized. HypoDD uses a mean-square-error data loss and explicit damping regularization. The code/method allows for using differential times from both cross-correlation and phase picks. At each iteration the algorithm checks for and discards observations based on criteria. In a typical use case, there is a user-defined maximum separation distance imposed between event pairs, which is progressively shrunk as the iterations continue. Each iteration requires inversion of a large sparse matrix containing the forward model. Uncertainties can be estimated by bootstrap resampling of the residuals and repeating the inversion multiple times. HypoDD uses all available observations in an inversion, provided they meet the criteria for each iteration. Practically speaking, datasets larger than a certain size need to be broken up into smaller subsets due to computational tractability.

\textbf{GrowClust}. The GrowClust method is a double-difference inversion scheme that uses cluster analysis to iteratively group events while relocating them \citep{trugman_growclust:_2017}. GrowClust defines a weighted similarity metric between event pairs based on cross-correlation coefficients and then sorts event pairs accordingly. GrowClust does not allow for the use of phase picks and uses only correlation data. Starting from the most similar event pairs, it merges clusters together and then relocates their centroids. GrowClust relies on robust loss functions, does not use explicit regularization, and optimizes the parameters with a grid search. The algorithm is designed to efficiently work through the dataset and only relocate as necessary upon successful merging of clusters. Since the method works sequentially through the dataset, there are limited opportunities to benefit from parallel processing; the main way that parallelization can be achieved is while bootstrap resampling the inversion results, where multiple inversions can be run in parallel. A key difference from HypoDD is that the hypocenters are not jointly optimized against the entire dataset. GrowClust explicitly constrains the centroid of the hypocenters to not move, and limits the grid search over the hypocenter changes to be within some pre-defined range.

\textbf{GraphDD}. This is a very recent method that aims to scale to very large seismicity catalogs \citep{mcbrearty_double_2025}. To achieve this, it incorporates graph neural networks to efficiently compute the forward model and double-difference objective. Graph neural networks are a very flexible and scalable deep learning architecture for learning on discrete meshes that are irregular, with variable number of nodes. GraphDD is a deterministic algorithm that minimizes a global L1-norm over the observations of the whole dataset jointly. In this regard it is more similar to HypoDD.
\section{Methods}

\subsection{Probability Model}
\noindent Inverse problems usually depend on many factors and assumptions, and it is important to quantify their combined effects on the solution.  In this regard, probabilistic inference is an appealing framework for solving inverse problems because the probability model combines all the assumptions together into one solution. From Bayes' rule \cite[e.g.]{tarantola_inverse_1982},
\begin{equation}
  p(\theta \mid \tau) \propto p(\tau \mid \theta)\, p(\theta).
\end{equation}
where $\tau$ are differential times and $\theta$ are the set of all hypocenter parameters. There are $M$ events, $L$ stations, and 2 phases, and the $n$th observation is associated with two sources $\xi_n^1,\xi_n^2$, one station $\eta_n$, and one phase $\phi_n$. We write the forward model for SPIDER as,
\begin{align}
    \tau_n &= \widehat{\tau_n}(\theta) + \Delta b_n + \epsilon_n\\
    \Delta b_n &= b_{\xi_n^1,\eta_n,\phi_n} - b_{\xi_n^2,\eta_n,\phi_n},
\end{align}
where $b_{m,\ell,\phi} \in \mathbb{R}$ is a latent event--station--phase term. This latent field captures correlated structure in the residuals that is not explained by the deterministic travel‐time model.

We collect residuals by station-phase group $g$ and write the residual vector as a sum of a shared‐event term and independent noise:
\begin{align}
  &\epsilon_n \equiv \tau_n - \widehat{\tau_n}(\theta) - \Delta b_n, \qquad
  r_g \equiv \{\epsilon_n\}_{n\in g}, \qquad \sigma_n \equiv \sigma_{\phi_n} \\
  &b_g \sim \mathcal{N}(0,\kappa_\phi^2 I), \qquad \epsilon_g \sim \mathcal{N}(0,\sigma_\phi^2 I)\\
  &r_g = W_g^{1/2} A_g b_g + \epsilon_g.
\end{align}
Here $A_g$ is the event‐pair incidence matrix for group $g$, and $W_g$ contains distance-based edge weights.

From eq (\ref{eq:fwd_model_error}) we see that the forward model error in the linearization regime is proportional to the absolute separation distance. Thus we use a diagonal edge‐weight matrix \(W_g\) to modulate the shared‑event term by event–event separation. Let the \(e\)th residual in group \(g\) correspond to event pair \((i_e,j_e)\) with distance \(d_e=\|x_{i_e}-x_{j_e}\|\). We define
\begin{equation}
  W_g = \mathrm{diag}(w_e), \qquad
  w_e = \left(\frac{d_e}{\ell}\right)^p + \varepsilon,
  \label{eq:correlation_decay}
\end{equation}
where \(\ell\) is a length scale, $p>0$ is a decay exponent, and \(\varepsilon>0\) prevents zero weights. $W_g$ is normalized such that the mean weight in each group $g$ is one.

We marginalize the latent variable:
\begin{align}
  p(r_g \mid \theta) &= \int p(r_g \mid b_g,\theta)\, p(b_g)\, db_g
  = \mathcal{N}(0,\Sigma_g),\\
  \Sigma_g &= \sigma_{\phi}^2 I + \kappa_{\phi}^2\, W_g^{1/2} A_g A_g^\top W_g^{1/2}.
\label{eq:marginal_sigma}
\end{align}
Thus the likelihood is a correlated Gaussian,
\begin{equation}
  p(\tau \mid \theta) \propto 
  \prod_{g}\exp\!\left(-\tfrac{1}{2}\, r_g^\top \Sigma_g^{-1} r_g\right)
  \prod_{n=1}^{N}\sigma_{\phi_n}^{-1}.
\end{equation}
The covariance matrix $\Sigma_g$ defined in eq. \ref{eq:marginal_sigma} contains two terms; the first of which is the typical IID noise and the second is a non-diagonal covariance matrix that results from the correlation between residuals with common events. Since covariance matrices are additive, properly accounting for this residual correlation inflates the variance for each observation and decreases the information content. Other double-difference relocation methods ignore this correlation and thus their uncertainties will be overconfident. 

For the prior $p(\theta)$ there are various reasonable options. One option is to use an uninformed prior and allow the data to entirely determine the posterior. However, there is often prior knowledge available to us that can be used to construct informative priors for our problem. The simplest is from our knowledge about the absolute uncertainty in hypocenters for the particular study area. For example, in California, it is believed that the absolute location errors for analyst reviewed earthquakes are generally less than about 500-1000m \citep{hauksson_waveform_2012} based on extensive validation against various data sources (i.e. quarry blasts). Similar reasoning is possible regarding the origin time. However, there are also instabilities in the inverse problem, which can arise, for example if all of the event pairs lie within a plane or surface, and such priors can help to address this instability. In a Bayesian setting, we can use this to construct a prior that limits the hypocenter shifts to something of this order, unless the data provide sufficient resolution / sensitivity.

We place a Gaussian prior on event parameters to regularize absolute location and origin‐time shifts:
\begin{align}
  &p(\theta_m) \sim \mathcal{N}(\mu_e,\,\Sigma_e), \qquad \mu_e \in \mathbb{R}^4,\\
  &\Sigma_e = \text{diag}(\sigma_{e,x}^2,\, \sigma_{e,y}^2,\, \sigma_{e,z}^2,\, \sigma_{e,t}^2).
\end{align}

The HypoDD method also considers priors that constrain the centroid of the seismicity due to separate instabilities of the inverse problem,
\begin{align}
       &p(\theta^c) \sim \mathcal{N}(\mu^c,\,\Sigma^c) \\
        &\mu_{c,j} = \frac{1}{M}\sum_m^M \theta_{mj}\\
        &\Sigma_c = \text{diag}(\sigma_{c,x}^2,\, \sigma_{c,y}^2,\, \sigma_{c,z}^2,\, \sigma_{c,t}^2)
\end{align}
This is particularly important because the double-difference forward model is relatively insensitive to systematic perturbations to a set of hypocenters, and the cloud of seismicity can start to drift because of the presence of observation noise. In standard inverse theory terms \citep{aster_parameter_2012}, centroid shift vectors are in the data null space. This problem is common when dealing with correlation data due to the relative sparsity of the connectivity matrix between event pairs.

\subsection{Stochastic Gradient Langevin Dynamics}
Our method for sampling from the posterior is based on a technique popularized in machine learning in recent years, called Stochastic Gradient Langevin Dynamics (SGLD) \citep{welling_bayesian_2011}. SGLD is a Markov chain Monte Carlo (MCMC) technique that incorporates stochastic gradients derived from mini-batches of data and balances these stochastic gradients against injected noise. To date, there has been very limited previous work in geophysics using SGLD \citep{siahkoohi_deep_2022}. In traditional stochastic gradient descent, parameter updates are generated iteratively as follows:
\begin{align}
    \Delta&\theta_t = \epsilon_t \nabla \log\hat{p}(\theta_t | y_t) \,\,\,= \,\epsilon_t \left(\nabla \log p(\theta_t) + \frac{N}{n}\sum_{i=1}^n \nabla \log p(y_{ti} | \theta_t)\right)
    \label{eq:sgd}
\end{align}
where $t$ is the current iteration index, $y_{ti}$ is a mini-batch of $n < N$ observations, and $\epsilon_t > 0$ is the learning rate. The terms in parentheses are a stochastic gradient approximation $\nabla \log\hat{p}(\theta_t | y_t)$ of the log posterior $\nabla \log p(\theta_t | y_{ti})$ using just $n$ observations instead of the whole $N$. Optimizing this equation results in the maximum a posteriori (MAP) solution.

In SGLD, the parameter updates are slightly modified by adding stochastic noise $(\sqrt{2\epsilon_t} \, \eta)$ to the stochastic gradients,
\begin{align}
    \Delta&\theta_t = \epsilon_t \nabla \log\hat{p}(\theta_t | y_t) + \sqrt{2\epsilon_t} \, \eta \,\,\,= \,\epsilon_t \left(\nabla \log p(\theta_t) + \frac{N}{n}\sum_{i=1}^n \nabla \log p(y_{ti} | \theta_t)\right) + \sqrt{2\epsilon_t} \, \eta  \\
    \eta& \sim \mathcal{N}(0,1) 
\end{align}
 Initially in the optimization, the stochastic gradients are large and dominate the parameter updates. As the algorithm converges toward a minimum, the stochastic gradients shrink and eventually become comparable in size to the injected noise. At this point, the algorithm naturally transitions from an optimization stage to a sampling stage, in which the stochastic gradients try to pull the parameters closer to the minimum, while the injected noise pushes them away. This is the essence of SGLD. 

Since its introduction, numerous subsequent studies have focused on improving SGLD \citep{kim_stochastic_2022,dubey_variance_2016,li_preconditioned_2015}. One notable development modifies SGLD to include a pre-conditioning matrix $V_t \in \mathbb{R}^{4M \times 4M}$ \citep{li_preconditioned_2015} (pSGLD), 
\begin{align}
    g_t &= \nabla \log \hat{p}(\theta_t | y_t) \\
    \quad v_t &= \beta_2 v_{t-1} + (1 - \beta_2) \, g_t \odot g_t, \\
    V_t &= \frac{1}{\sqrt{v_t} + \epsilon}\\
        \theta_{t+1} &= \theta_t + \epsilon_t V_t \odot g_t + \sqrt{2\epsilon_t V_t} \odot \eta,
    \label{eq:pSGLD}
\end{align}
where $\beta_2 \in [0, 1)$ and $\odot$ denotes elementwise multiplication. The pre-conditioning matrix adjusts the variance of the injected noise for each parameter based on the magnitude of the gradient, helping to improve convergence in situations where the parameter spaces are relatively imbalanced. Another variant is Stochastic Gradient Hamiltonian Monte Carlo (SGHMC) \citep{chen_stochastic_2014}, which tries to combine ideas from Hamiltonian Monte Carlo, with the scalability and flexibility of stochastic gradients. In this paper we focus on the pSGLD algorithm but note the potential for using SGHMC as well.

We take $\theta_0$ as the initial absolute locations provided by a single-event hypocenter inversion code. Parameter updates are performed on mini-batches of $n$ observations and there are $N/n$ mini-batches per epoch. We split the inversion up into four distinct stages to maximize optimization efficiency, and stabilize the algorithm to relatively similar hyperparameters across four catalogs. In the first stage, we obtain a robust initial solution using eq. \ref{eq:stage1loss} with a L1 norm and the Adam optimizer with $\eta=\num{1e-3}$. We then discard outlier observations that have residuals greater than 4 median absolute deviations. In the second stage, we warm up a pSGLD sampler using eq. \ref{eq:pSGLD} starting from the Stage 1 solution for a fixed number of epochs, with the injected noise turned off. In the third stage, we linearly ramp up the injected noise from zero to full amplitude over a fixed number of epochs. Finally, in the fourth stage, we iteratively collect samples from the posterior. In computing the parameter updates, we evaluate the forward model as in eq. \ref{eq:forwardmodel}. We use EikoNet as the forward model \citep{smith_eikonet_2020}, a pre-trained physics-informed neural network Eikonal solver that is differentiable. The gradients $\nabla \log\hat{p}(\theta_t | y_t)$ are computed with reverse-mode automatic differentiation.

Fundamentally, SGLD is an MCMC algorithm and is based on collecting samples. However, a major difference from other MCMC algorithms is that SGLD is most commonly used without a Metropolis-Hastings rejection step, i.e. samples are always collected at each iteration. \citet{welling_bayesian_2011} show that this is possible because when $\epsilon_t$ is small, the rejection probability becomes negligible. In practice, SGLD often uses a fixed $\epsilon$ that is tuned to be just small enough to satisfy these conditions (more details in a later section). As with other MCMC algorithms, the samples will be highly correlated and usually benefit from thinning. An additional parameter that controls convergence is the batch size, which is not present in typical MCMC algorithms that use the entire dataset for each parameter update. We provide guidance on the choice of learning rate and batch size in a subsequent section.

\subsection{On the Use of Differential Times from Phase Picks}
Let $\hat{t}_i = t_i + \epsilon$ be a noisy measurement of the arrival time $t_i$, where $\epsilon \sim \mathcal{N}(0,\sigma^2) $.  In some instances \citep{waldhauser_double-difference_2000}, it has been advocated to form differential times from these noisy phase picks, i.e.,
\begin{equation}
    \hat{\tau}_{ij} = \hat{t}_i - \hat{t}_j
\end{equation}
We will show that this operation couples the observations together, making $\hat{\tau}_{ij}$ correlated. Suppose that, given a vector of observations $\hat{t} \in \mathbb{R}^N$, we form $\hat{\tau}_{ij} \; \forall \, i<j $, i.e. between all pairs of events. In this case, $\hat{\tau} \in \mathbb{R}^K$, with $K = \frac{N!}{4(N-2)!}$. With $k = (i,j)$, we can define the differencing matrix
\[ V_{kl} = \begin{cases} 
      1 & l=i \\
      -1 & l=j \\
      0 & \text{otherwise},
   \end{cases}
\]
where $V\in\mathbb{R}^{K\times N}$. Then, the differential times $\hat{\tau}$ can be formed as
\begin{equation}
    \hat{\tau} = V \hat{t}.
\end{equation}
Since the $\hat{t}_i$ are independent and identically distributed Gaussian random variables with variance $\sigma^2$,  then,
\begin{equation}
\text{cov} \,(\hat{t}) = \sigma^2 I    
\end{equation}
However, 
\begin{equation}
    \text{cov}  (\hat{\tau}) =  \text{cov} (V \hat{t}) = V  \,\text{cov} (\hat{t})\, V^\top =  V (\sigma^2 I) V^\top = \sigma^2 V V^\top,
\end{equation}
where $ V V^\top$ in general is a non-diagonal matrix. Thus, if one forms differential times from phase picks, it is essential to include the full covariance matrix $\Sigma = \sigma^2 V V^{\top}$ in the likelihood  $p(\tau |\,\theta, \eta)$ for the uncertainty estimates to be correctly inferred. This correlation structure is not accounted for in HypoDD and therefore bootstrap estimates based on differential phase picks will not be reliable. This is not relevant for GrowClust because it does not explicitly allow for differential phase picks. Dealing with the correlated error injected into the residuals is non-trivial because it mixes with the forward model error, and as such, we have designed the method strictly for differential times from cross-correlation.

\subsection{Uncertainty Quantification}
Assuming that the full posterior for millions of parameters is able to be determined, an immediate question is how such a solution can be used in a way that properly takes advantage of all this information. The simplest description of the hypocenters individually comes from determining scalar credibility intervals. A non-parametric way to do this is take the MCMC samples and compute the (half) length of the credibility interval defined by two given percentiles, e.g. [0.5, 99.5] percentile, such that the interval contains 99\% of the probability. For any one event, these Cartesian-type uncertainties provide upper bounds to the uncertainties that are independent of correlation structure. It is equivalent to placing a 3D box centered on the median solution and does not assume that the posterior is a multivariate Gaussian.

The full posterior can be used to study the catalog-wide uncertainties and correlations. Specifically, we can stack all of the X samples, Y samples, Z samples, and origin time samples, which is equivalent to marginalizing over the catalog. These marginalized distributions represent the properties of a typical event in the catalog. The resulting covariances can be visualized to obtain catalog-wide snapshots of the relationships between the hypocentral parameters that reflect the network geometry, measurement error, and other data conditions.

It is also possible to use the full posterior to bin all of the spatial samples in 3D. Alternatively one could use kernel density estimation (KDE). This allows for plotting of the KDE in 2D slices to interpret whether certain geometric features in the seismicity are reliable. Since the posterior represents the full uncertainty associated with each event, the KDE therefore will blur out features that are smaller than the uncertainties locally. As we will show in the next section, it allows for spatial estimation of quantities such as fault zone width.

\subsection{Parallelism and Scaling}
SPIDER is highly parallelizable and designed for GPUs. In the subsequent experiments, we use one or two Nvidia RTX 6000 Ada GPUs, depending on the dataset. It is trivial to use multiple with data parallelism, i.e. splitting the mini-batches over multiple GPUs when computing gradients. Generally this comes along with increasing the batch size to compensate for the extra GPU devices. Another possibility is to run multiple sampling chains over the multiple GPUs, instead of splitting up the batch, to achieve essentially linear scaling.

\subsection{Guidance on the choice of learning rate and batch size}
The 2-Wasserstein distance between the generated samples of the discrete dynamics and the continuous dynamics is $\sqrt{\eta^{\frac{1}{4}}}$ \citep{raginsky_non-convex_2017}. Therefore, we need to pick the learning rate small enough that it recovers a distribution that is "close" to the true distribution. At the same time, we need to pick the learning rate large enough that the corresponding continuous dynamics mixes quickly into its stationary Gibbs distribution. This is exponential, i.e., $\exp{(-c\eta)}$, with $c$ a large coefficient describing the spectral gap \citep{raginsky_non-convex_2017}. From the perspective of deep learning, it is necessary to pick the learning rate to be large enough that the dynamics can jump over the narrow valleys of the likelihood loss landscape. In the end, when tuning the learning rate for the problem at hand, one needs to consider the interplay between these components. A learning rate too large gives a meaningless posterior, due to the first reason, whereas a learning rate too small makes our model mix very slowly, making the optimization inefficient. A valuable metric for tuning the learning rate is the variance of the stochastic gradient to the variance of the Langevin injected noise. From theory it can be shown that this ratio depends linearly on the learning rate, and a value no higher than $0.1-1.0$ will generally get the sampler into a good range.

Regarding the batch size, there are additional tradeoffs. Larger batches allow for greater parallelization over GPUs, leading to faster run time per epoch; however we found empirically tests that with Nvidia RTX 6000 Ada GPUs, the effective upper limit is roughly 1-2 million differential times per GPU. Of course, this limit will also depend on the number of events in the catalog. We found that for all tests in this study the problem was GPU compute bound and not memory bound. In terms of convergence of the SGLD sampler to the stationary distribution, larger batch sizes lead to stochastic gradients that have smaller variance, which can also generally help to converge more quickly \citep{welling_bayesian_2011,dubey_variance_2016}. However, for the same reason, it may also make the optimization phase of SGLD slower, and therefore take longer to transition to the sampling phase. However, when the batch size is $>1\%$ of the total dataset, this stochastic gradient variance is going to be relatively small.

\subsection{Datasets}

\noindent\textit{Synthetic Benchmark Catalog for 2019 Ridgecrest sequence}\\
This synthetic dataset was created by \citet{yu_accuracy_2024} for the purpose of benchmarking different hypocenter location algorithms. It consists of an earthquake catalog and synthetic phase picks. The dataset contains 1000 earthquakes with "true" locations taken from the 2019 Ridgecrest, California sequence. A realistic synthetic 3D velocity model with power law heterogeneity was used to generate observations. From the raw arrival times without noise added, we form differential times for all event pairs within 1 km separation distance. We then add 10 msec Gaussian IID noise to each differential time.\\

\noindent\textit{Cahuilla Swarm in Southern California}\\
This dataset covers the Cahuilla earthquake swarm in southern California \citep{ross_3d_2020} over the period 2016-2020. It contains 22,701 earthquakes and about 76M differential times. Nearly all events in the Cahuilla swarm occur within a single spatial cluster about 3 km in size.\\ 

\noindent\textit{Ridgecrest, California Sequence}\\
This catalog covers the 2019-2023 Ridgecrest earthquake sequence in southern California and was produced by \citet{atterholt_evolution_2025}. The dataset was produced with a deep learning phase picker called Phase Neural Operator \citep{sun_phase_2023}, GaMMA phase associator \citep{zhu_earthquake_2022}, and HypoSVI for absolute locations \citep{smith_hyposvi_2021}. The catalog is by far the largest tested in this study, with 326,346 events that are associated with 30M differential times. The events also cover the largest spatial area, roughly 75 km by 75 km, with the network itself having a much larger footprint. The seismicity forms numerous disjoint clusters and thus is not connected into essentially a single graph like the Cahuilla Swarm is. \\

\noindent\textit{Noto, Japan Sequence}\\
This dataset encompasses the 2020–2024 Noto, Japan earthquake sequence, which is notable for several years of swarm-like seismicity that ultimately culminated in an $M_w$ 7.5 mainshock in January 2024 \citep{kato2024implications,peng_evolution_2025}. We analyzed continuous waveform data recorded at 48 permanent seismic stations in and around the target region, covering the period from 1 February 2023 up to the mainshock rupture. To identify P- and S-wave arrival times, we applied PhaseNet, a deep neural network-based picker \citep{zhu_phasenet:_2019}. After extracting the first-arrival time series, we associated the phase picks with individual events using REAL algorithm \citep{zhang_rapid_2019} and a one-dimensional velocity structure \citep{1370004229808520843}. The initial catalog has 72669 events. Differential arrival times were computed using a waveform cross-correlation method. Each P- and S-waveform was bandpass-filtered between 2–12 Hz and cross-correlated over 1.0 s time windows \citep{kato_conjugate_2021}

\section{Results}
In this section we discuss inversion results for each dataset. All catalogs used $\beta=0.99$, $\epsilon=\num{1e-5}$, $\ell=5.0$km, and outlier removal after Stage 1 using $4\times$ the median absolute deviation of the residuals. The event prior is the same for the three real catalogs, with $\sigma_{e,x}=\sigma_{e,y}=\sigma_{e,z}=0.2$ km and $\sigma_{e,t}=0.1$ sec, and the centroid prior is the same for all four catalogs, with $\sigma_{c,x}=\sigma_{c,y}=\sigma_{c,z}=0.01$ km and $\sigma_{c,t}=0.01$ sec. The remaining tuned hyperparameters for each catalog are given in Table \ref{tab:algorithm_parameters}. For SPIDER we ensure that the observations for inversion are in the linearization regime by removing those with linearization error greater than 5\%; intuitively, this means that the difference between the non-linear forward model and linearized forward model is no greater than 5\% for any observation.
\begin{table}[h]
    \centering
    \fontsize{8pt}{11pt}\selectfont
    \setlength{\tabcolsep}{4pt} 
    \resizebox{\linewidth}{!}{%
    \begin{tabular}{
        l
        *{4}{S[table-format=5.0]}
        S[scientific-notation = true, table-format=1.1e1] 
        S[scientific-notation = true, table-format=1.1e1] 
        S[scientific-notation = true, table-format=1e1] 
        S[table-format=1.3]
        l
        l
        l 
    }
        \toprule
        \textbf{Catalog} & \multicolumn{4}{c}{\textbf{Stage}} & {$\eta$} & \textbf{Batch Size} & {$\sigma_p=\sigma_s$ (s)} & {$\kappa_p$ (s)} & {$\kappa_s$ (s)} & {$p$} & $\ell$ (km)\\
        \cmidrule(lr){2-5}
         & {1} & {2} & {3} & {4} &  &  &  &  &  &  & \\ 
        \midrule
        Ridgecrest (Synthetic) & 1000 & 100 & 100 & 100000 & 1e-2 & N/A & 2e-2 & 0.033 & 0.06 & 0.25 & 40.0\\
        Cahuilla Swarm         & 150 & 100 & 100 & 15000  & 1e-3 & 1.5e6 & 1e-2 & 0.03 & 0.04 & 0.25 & 40.0\\
        Ridgecrest (Real)      & 2000 & 100 & 100 & 15000   & 1e-2 & 1e6 & 1e-2 & 0.03 & 0.04 & 0.25 & 40.0\\
        Noto Swarm             & 2000 & 100 & 500 & 15000 & 1e-4 & 1e6 & 1e-2 & 0.03 & 0.04 & 0.25 & 40.0 \\
        \bottomrule
    \end{tabular}%
    }
    \caption{Algorithm hyperparameters for the catalogs in this study.}
    \label{tab:algorithm_parameters}
\end{table}
\subsection{Synthetic Benchmark Catalog for 2019 Ridgecrest sequence}
\begin{figure}[h!]
    \centering
    \includegraphics[width=0.95\linewidth]{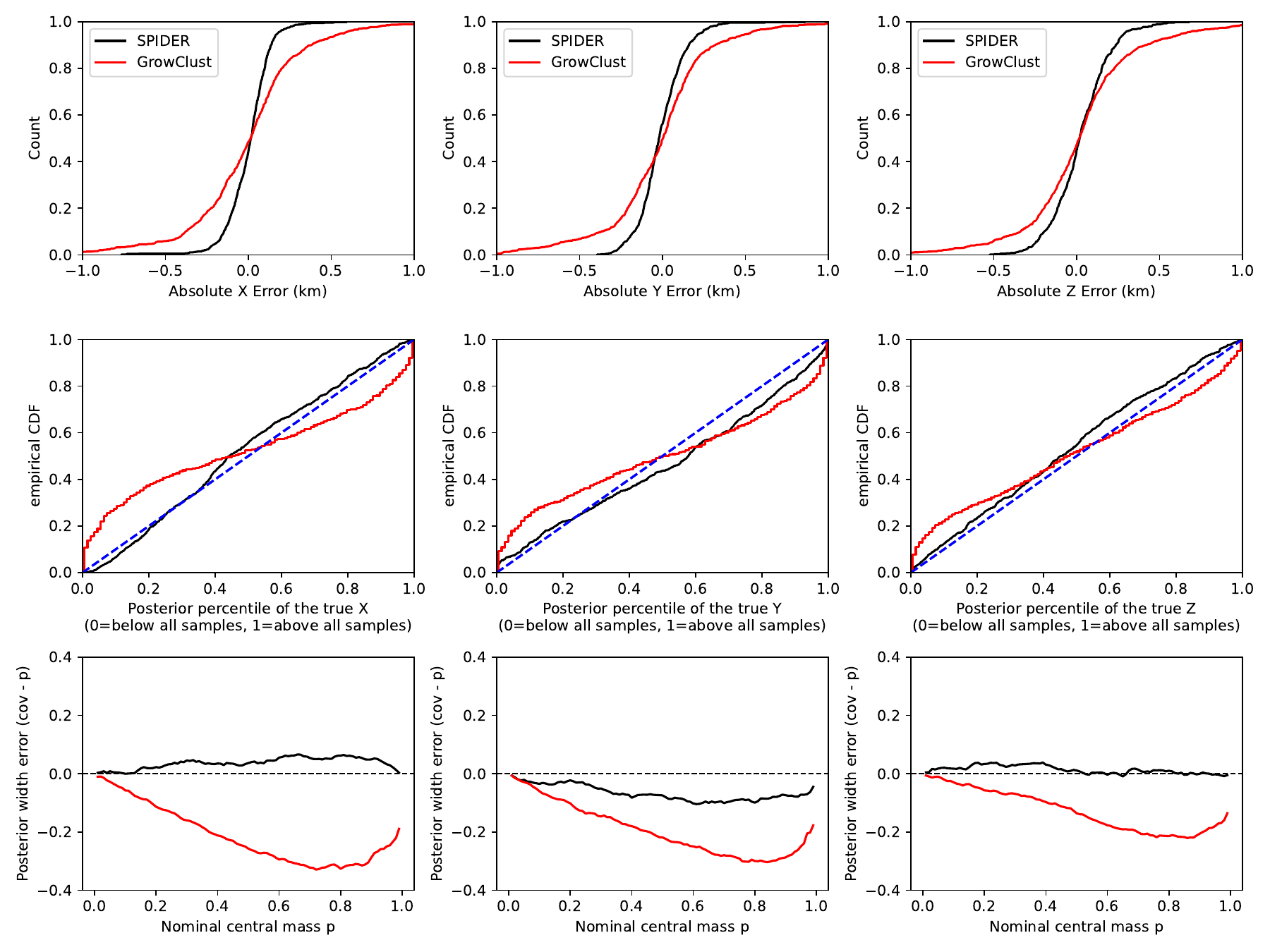}
    \caption{Location errors and uncertainties for the synthetic Ridgecrest catalog. Upper row contains absolute errors for the posterior mean compared with GrowClust best-fit. Middle row shows marginal posterior calibration plots for uncertainty quantification (see definition in main text). Lower row shows signed posterior width error vs central mass percentile. Negative values indicate the posterior is too narrow, whereas positive values indicate the posterior is too wide.}
    \label{fig:syn-error}
\end{figure}

The 1D velocity model given by \citet{yu_accuracy_2024} was used. Since the synthetics were generated from a velocity model with complex 3D structure, there is significant mismatch between the 1D forward model predictions and the observations, even at the true locations. We thinned the samples by a factor of 10. We require a minimum of 6 differential times per event pair. The event prior uses the initial locations as the mean with $\sigma=0.5$ km for the spatial coordinates and $\sigma=0.1$ s for the origin time. We draw initial locations from this prior. All additional parameters are given in Table \ref{tab:algorithm_parameters}. The exact absolute errors for the mean posterior are shown in Fig. \ref{fig:syn-error} for both SPIDER and GrowClust, which are in good agreement for this synthetic catalog. SPIDER's mean locations rarely exceed 250 m error, and GrowClust is slightly worse.

In Fig. \ref{fig:syn-error} we show uncertainty calibration plots for both methods, which evaluate how good the posterior represents the true error. For this test, we used 10,000 SPIDER posterior samples and 100 GrowClust bootstrap samples. Then for each event and each coordinate, we evaluate where the true value falls among the posterior samples, which is a percentile $\in [0,1]$. If the posterior is perfectly calibrated, these percentile values will be uniformly distributed. Deviations from the diagonal can inform whether the posterior is under- or over-dispersed across the dataset. Finally, we can also compute the signed posterior width error for a set of central mass percentiles. This quantity tells us whether the posterior is too narrow (negative) or too wide (positive). For the X coordinate, SPIDER's posterior is about 5\% too wide systematically, whereas GrowClust is systematically too narrow, being about 35\% small in the tails. For the Y coordinate, SPIDER's posterior is slightly too narrow (at most 8\%), whereas GrowClust is systematically 20-30\% too narrow. Finally for the Z coordinate, SPIDER's posterior is very reliable, only 3-5\% too wide at most, whereas GrowClust is about 10-20\% too narrow. From this experiment we conclude that SPIDER is effective and the uncertainty estimates are geophysically meaningful. 

\subsection{Cahuilla Swarm}

\begin{figure}[htbp!] 
    \centering
    \includegraphics[width=0.9\linewidth]{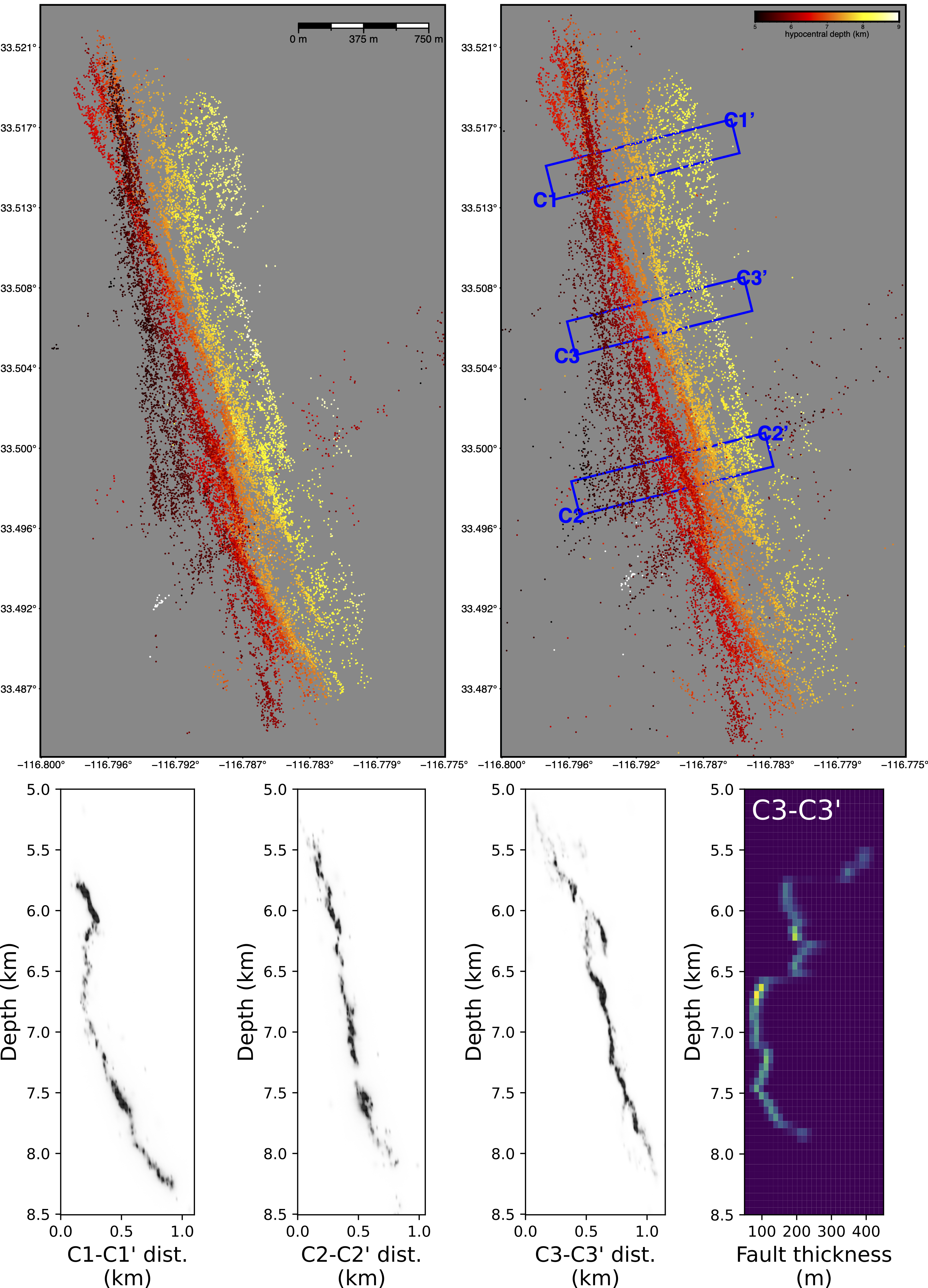}
    \caption{Comprehensive analysis of the Cahuilla Swarm. Top: Comparison of relocation methods, with GrowClust on left and SPIDER on right. Bottom: Posterior density and cross-sections C1-C3.}
    \label{fig:cahuilla_combined}
\end{figure}

\begin{figure}
    \centering
    \includegraphics[width=0.95\linewidth]{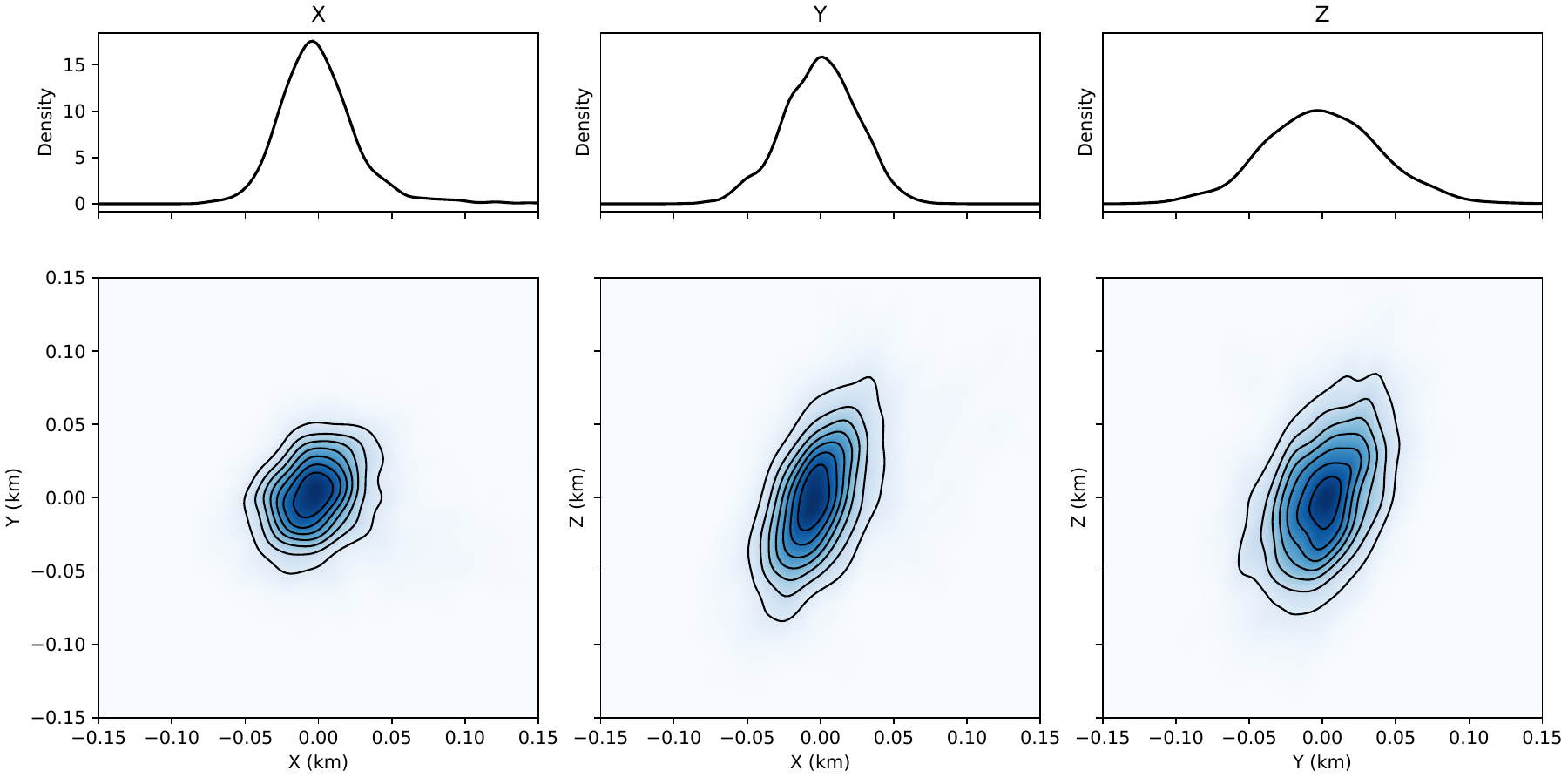}
    \caption{Cahuilla swarm catalog-wide posterior densities stacked over all events. Upper row shows 1D marginal posteriors for X, Y, and Z, whereas lower row shows 2D joint posteriors. These diagrams give a snapshot of the uncertainties and correlations (if present) that persist over the entire catalog.}
    \label{fig:cahuilla_marginals}
\end{figure}

For the inversion, we used the 1D velocity model of \citet{hadley_seismic_1977}. We use a minimum of 6 differential times per event pair and a minimum cross-correlation coefficient of 0.6. All other parameters are given in Table \ref{tab:algorithm_parameters}. In total, 20,954 events are relocated by GrowClust and 22,143 events are relocated by SPIDER. Figure \ref{fig:cahuilla_combined} shows a close-up of the locations, indicating that they are overall very similar between the two methods for this dataset. The spatial compactness of the swarm results in events that are very well connected in terms of differential times, with high correlation coefficients in general ($>0.8$).
 
The general uncertainty characteristics of this catalog are described in Figure \ref{fig:cahuilla_marginals}, in which the samples for all events in the catalog are stacked. The upper row of this figure shows the marginal distributions for the three hypocentral coordinates, whereas the lower row shows joint posterior distributions between the three unique pairs. This figure is useful to qualitatively assess the large-scale properties of the catalog in terms of uncertainties and any correlations that might be present. It should be interpreted as the posterior of a typical event in the catalog, which show that the typical event exhibits moderate tradeoff between the two horizontal directions, but that its depth is more strongly correlated with the horizontals. Also shown are  per event based on the 99\% credibility intervals. These events are very tightly constrained both horizontally and vertically, with a median horizontal uncertainty of about 8 meters and a median vertical uncertainty of about 14 meters. 

We plot the posterior mean catalog for the Cahuilla dataset in Figure \ref{fig:cahuilla_combined} together with along-dip depth sections that illustrate the relative density of all stacked posteriors in the catalog along narrow fault sections. Visualizing the entire chain enables us to examine interpreted structural features of the catalog and assess whether these features are significantly constrained relative to location uncertainty. This exercise provides corroborating evidence for several interpretations made in \citep{ross_3d_2020}. First, the sharp concentration of sample points enables us to confirm that the along-strike and along-dip variations in planarity identified in \citep{ross_3d_2020} are robust observations. Furthermore, along-dip variations in posterior density substantiate the horizontal fault channeling originally interpreted in \citep{ross_3d_2020}.

In their assessment of the GrowClust locations initially calculated for this catalog, \citep{ross_3d_2020} suggests that they constrain the seismogenic thickness of the Cahuilla fault to be on the order of tens of meters. Our analysis allows us to probabilistically interrogate this estimate. To demonstrate, we subset the catalog to earthquakes within the fault cross-section C3-C3’. For each realization of the subset catalog, we estimate fault thickness as a function of depth. We begin by sorting events by depth and moving a sliding 200-event window over the depth range of the realization’s hypocenters. Within each window, we perform a linear regression to the windowed events’ fault-perpendicular (C3-C3’ direction) and depth coordinates; we estimate fault thickness as the difference between the 2.5th and 97.5th percentiles of the regression residuals. This process returns a simplified estimate of seismogenic thickness in the depth window. We then stack the resulting thickness measurements for all realizations of the catalog to obtain a distribution over measured fault thicknesses (Fig. \ref{fig:cahuilla_combined}). For the C3-C3’ events, we constrain a maximum fault thickness of approximately 75 m at 7.0 km depth. This more precise estimate is comparable to the uncertainty estimates derived from the catalog’s posteriors, suggesting that this value may indeed be an upper bound.

\subsection{Ridgecrest, California Sequence}
\begin{figure}[htb!]
        \centering
        \includegraphics[width=0.95\textwidth,trim={0cm 0cm 0cm 0cm},clip]{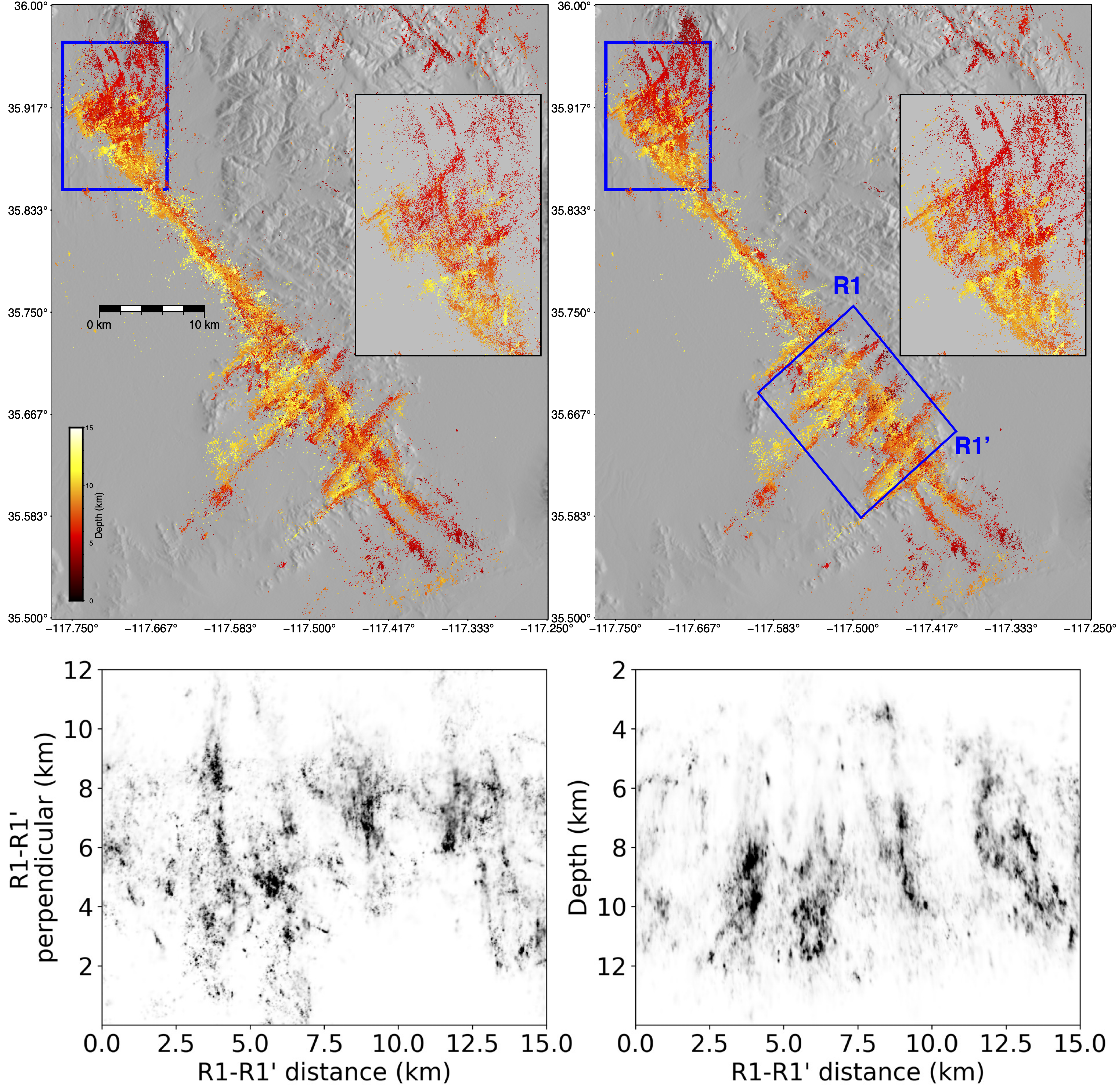}
        \caption{Ridgecrest real-data catalog relocation results. Upper left: GrowClust baseline. Upper right: SPIDER. Insets: closer view of locations within the black box. Lower left: stacked posterior densities for the R1-R1' section in map view, highlighting fault strand details. Lower right: stacked posterior densities for R1-R1' in depth view, highlighting the separation and dip of individual fault strands.}
        \label{fig:ridgecrest_mapview}
\end{figure}

We use a minimum of 6 differential times with cross-correlation coefficient of 0.6 or greater and inversion parameters as defined in Table \ref{tab:algorithm_parameters}. After relocating the events with SPIDER, we obtain 237,203 solutions, while GrowClust results in 214,467 solutions. Figure \ref{fig:ridgecrest_mapview} shows the seismicity in map view and cross-section compared to GrowClust. In both cases, the locations are broadly consistent with the catalog from \citet{ross_hierarchical_2019}, which covered just the first 21 days of the sequence. There are some differences noticeable between the two catalogs; namely in the SPIDER results there is less of a tendency to split continuous seismicity features into multiple disjoint clusters. This is most noticeable in the seismicity at the very northwest edge of the rupture, where the horse-tail faults exhibit lateral offsets of about 500 meters in the GrowClust results, whereas the SPIDER results show these as continuous. Overall, however there are no major differences.   

We also provide density plots representing all catalog samples for section R1-R1' in Figure \ref{fig:ridgecrest_mapview} (bottom). Similar to the Cahuilla sample plots, these stacked-posterior plots demonstrate that the lengthscales of individual event posteriors are far smaller than the lengthscales of resolvable structures in the catalog. The map view of sample density highlights the predominance of NE-striking fault strands along this segment of the Ridgecrest fault system. The depth view reveals the depth extent of these strands and illustrates the persistence of complex, multiscale fault structures from the surface to mid-crustal ($\sim12$ km) depth.

\begin{figure}[htb!]
    \centering
    \includegraphics[width=0.95\linewidth]{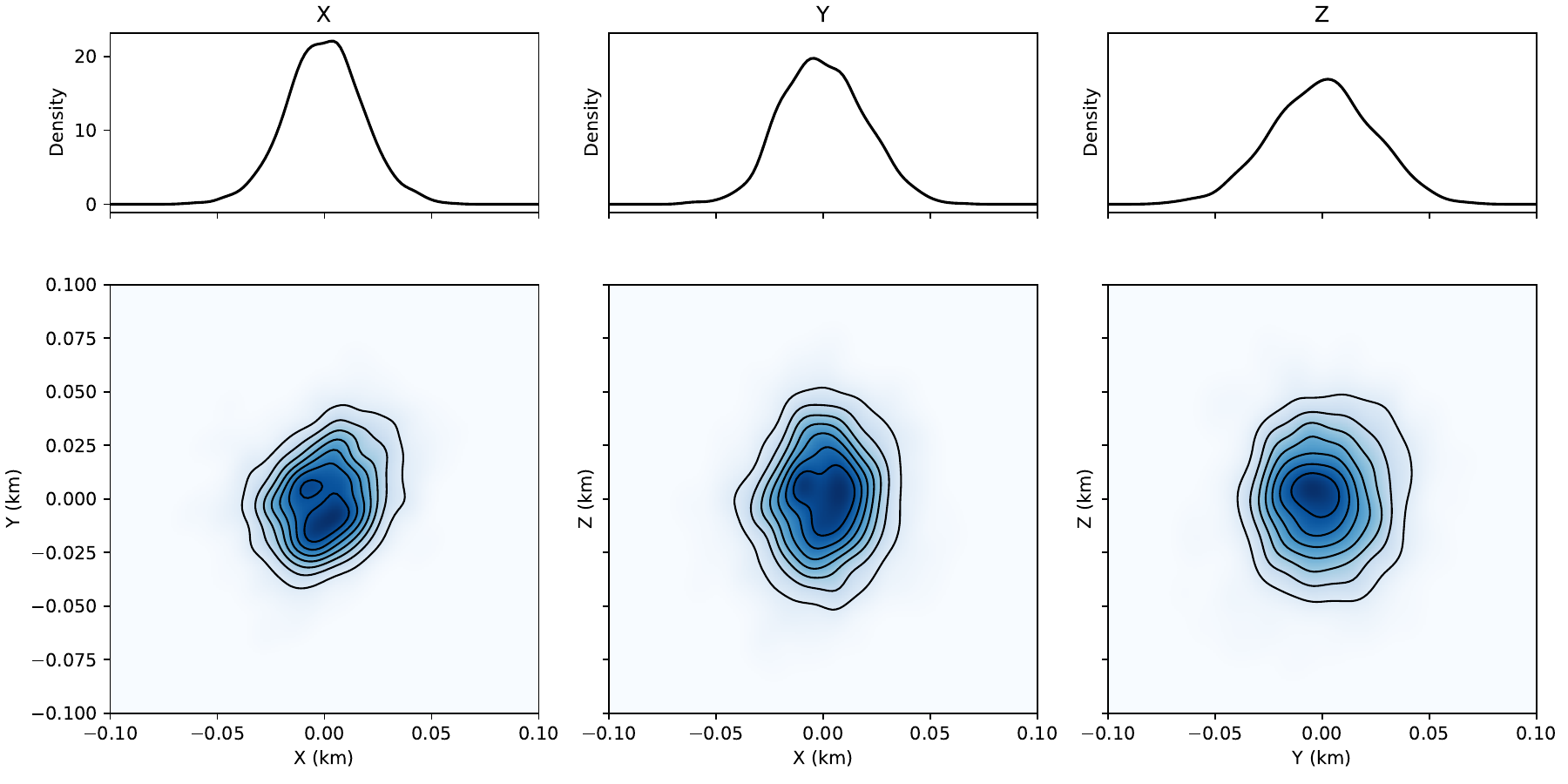}
    \caption{Ridgecrest (real data) catalog-wide posterior densities stacked over all events. Upper row shows 1D marginal posteriors for X, Y, and Z, whereas lower row shows 2D joint posteriors with contours. These diagrams give a snapshot of the uncertainties and correlations that persist over the entire catalog.}
    \label{fig:ridgecrest_marginals}
\end{figure}

Figure \ref{fig:ridgecrest_marginals} shows the general uncertainty characteristics of this catalog, in which the samples for all events in the catalog are stacked. From this we conclude there are no systematic correlations between the three hypocentral coordinates across the catalog and that the horizontal uncertainties are not biased toward any particular azimuthal direction. There may be very weak correlation between X-Y.

\subsection{Noto, Japan Earthquake Sequence}
\begin{figure*}[h!]
  \centering
    \includegraphics[width=1.0\linewidth]{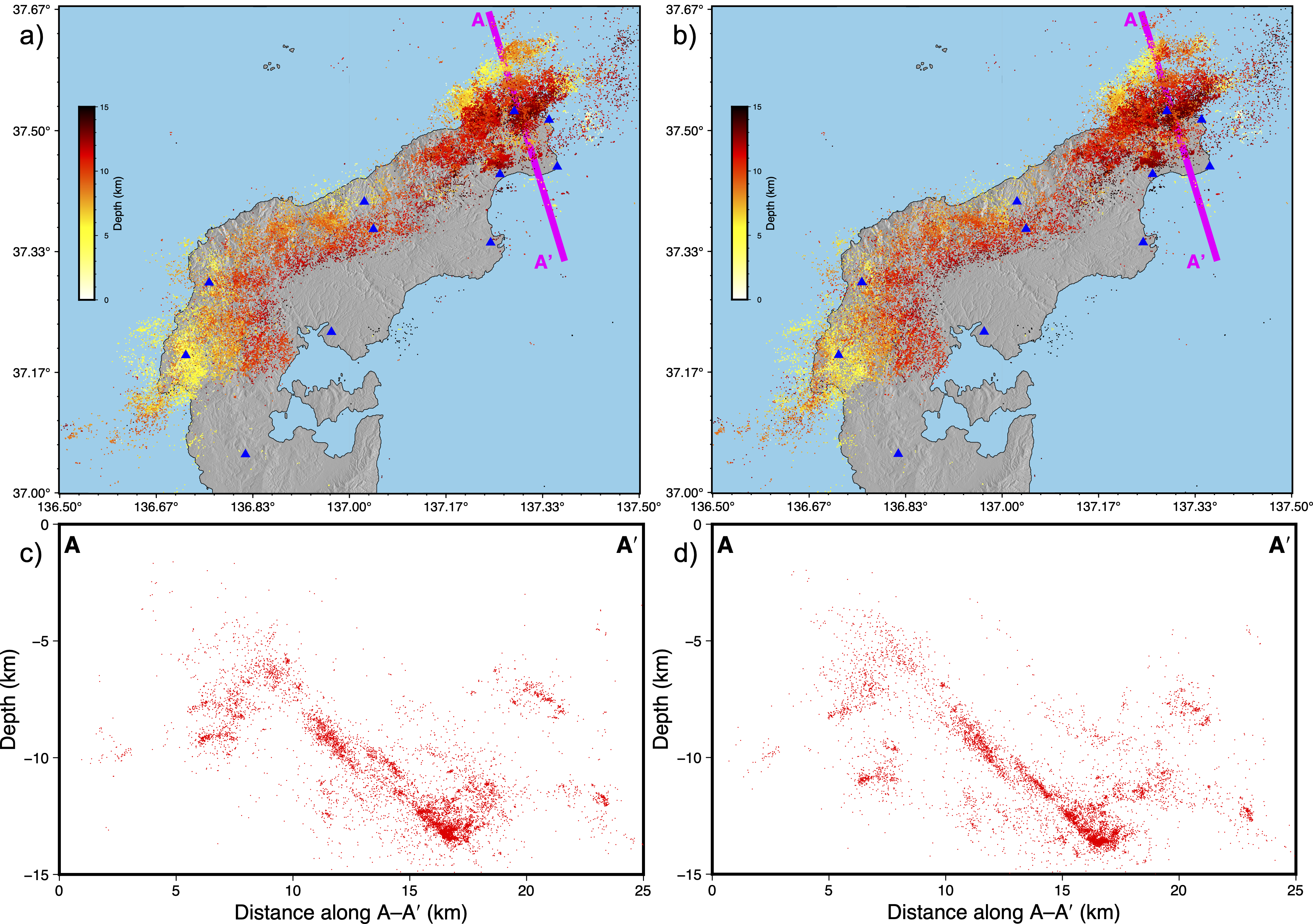}
  \caption{Noto swarm sequence relocated seismicity catalogs. (a,c) GrowClust; (b,d) SPIDER. Events shown on cross section A–A' are within 2.5~km of the magenta line}
  \label{fig:noto_seismicity}
\end{figure*}

For the inversion we use a minimum of 4 differential times with cross-correlation coefficient of 0.8 or greater. The remaining inversion parameters are given in Table \ref{tab:algorithm_parameters}. There are a total of 56,989 events that meet the minimum criteria for these settings in the SPIDER catalog and 59,987 in the GrowClust catalog. The mean of the posterior is shown in Figure \ref{fig:noto_seismicity}. The SPIDER results are slightly sharper and more compact. The main fault structure seen in the cross-sections has a lateral offset between the deep and shallow parts, likely due to sparse offshore station coverage. This offset is more pronounced in the GrowClust locations compared to those obtained with SPIDER. 

\begin{figure}[htb]
    \centering
    \includegraphics[width=0.95\linewidth]{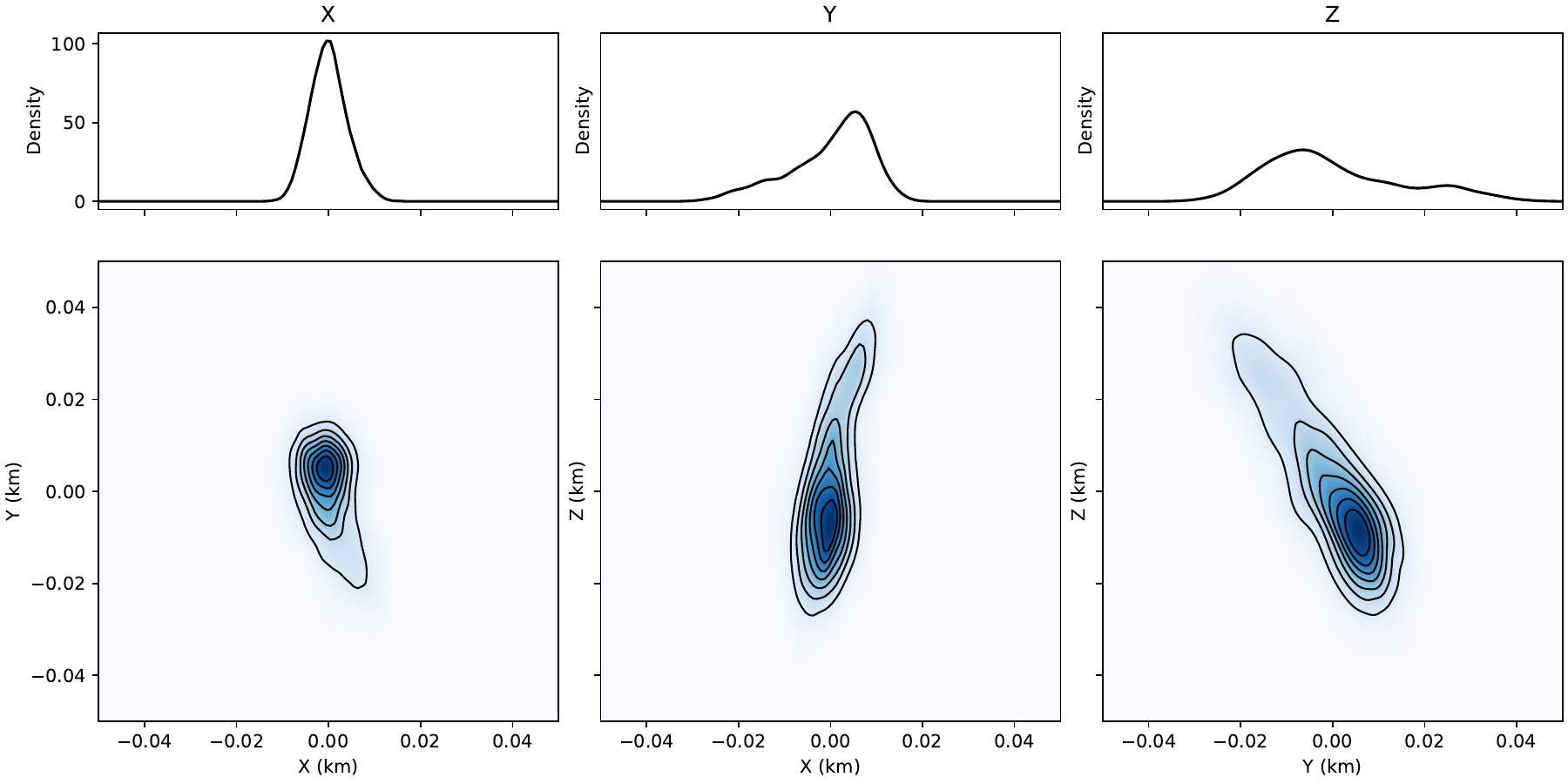}
    \caption{Noto catalog-wide posterior densities stacked over all events. Upper row shows 1D marginal posteriors for X, Y, and Z, whereas lower row shows 2D joint posteriors with contours. These diagrams give a snapshot of the uncertainties and correlations that persist over the entire catalog.}
    \label{fig:noto_marginals}
\end{figure}

Figure \ref{fig:noto_marginals} shows the catalog-wide posterior densities stacked over all events. We can see that in this case the joint densities are strongly non-Gaussian with heavy tails, reflecting the poor array geometry and a broad range of location qualities across the dataset.

\subsection{Quantifying shared-event residual correlation}
\begin{figure}[h!]
    \centering
    \includegraphics[width=0.9\linewidth]{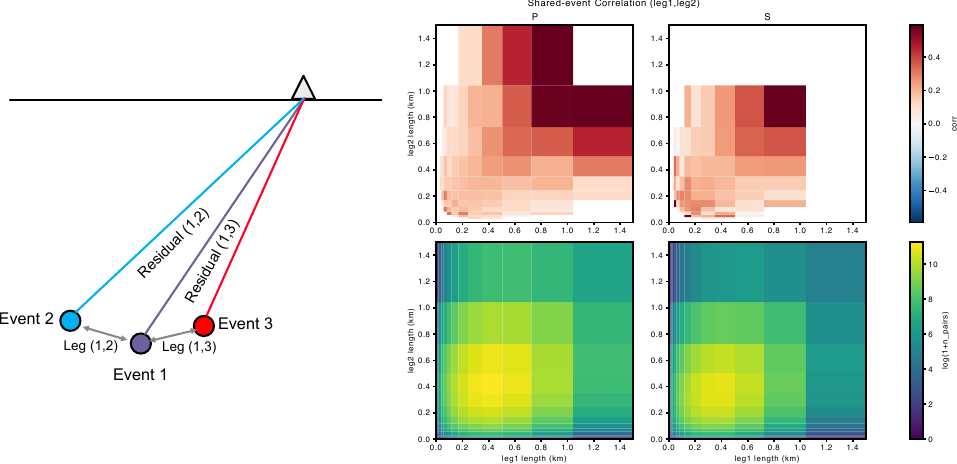}
    \caption{Demonstration of distance-dependent residual correlation between pairs of differential times with a common event.}
    \label{fig:shared_error}
\end{figure}

We finally demonstrate the severity of the residual correlation and the need for properly accounting for it. For the Cahuilla Swarm dataset, we compute the forward model for the best-fitting locations at the end of Stage 1, and construct residuals. We focus on pairs of residuals for which the two differential times have a common event at the same station and phase (see diagram in Fig. \ref{fig:shared_error}). We refer to the event separation for residual 1 as leg 1 length, and the event separation for residual 2 as leg 2 length. Figure \ref{fig:shared_error} shows that the correlation is near zero when the separation is close to zero, and increases with separation distance to a maximum correlation value of about 0.5 when both leg lengths are about 1 km or greater. Similar plots are observed for all other datasets, both at the starting locations, and at the MAP locations after relocation. This justifies the need for the extra model complexity in SPIDER.

\section{Discussion}

We have presented a new method, SPIDER, which allows for efficient Bayesian inference of earthquake hypocenters with double-difference methods. The method was designed with a particular emphasis on scalability, as seismicity catalogs are routinely reaching hundreds of thousands to millions of events at present. Since the problem is formulated as one of Bayesian inference, prior information about the hypocenters, and the mathematical structure of the observation noise can be incorporated to yield posterior distributions that more accurately reflect uncertainties in the inversion results. To achieve this, we use a combination of Stochastic Gradient Langevin Dynamics with reverse-mode automatic differentiation to efficiently scale up to massive datasets while sampling the posterior distribution. We applied the method to one synthetic and three real catalogs to demonstrate the results under a wide range of conditions. Our testing suggests that SPIDER's mean or MAP locations vary only slightly from other double-difference relocation methods and these differences are effectively invisible by eye; thus our emphasis is on scalability and reliable uncertainty quantification.

There are many approaches to Bayesian inference and in particular there have been numerous advances in the last decade or so that have allowed for greater scalability to large datasets, many of which have been motivated by uncertainty quantification in machine learning. Variational inference, which presents a scalable alternative to SGLD, casts the problem of posterior estimation as an optimization problem over a flexible family of distributions. Other types of posterior estimation methods include normalizing flow networks, and diffusion models. Each method has advantages and disadvantages.

The choices for distributions discussed in this paper are not intended to be set in stone. We intend to expand options for distributions implemented in SPIDER within mathematical and computational limitations. The goal is to let the user decide what they think are the right distributions for the problem, which we believe is a key part of Bayesian inference. 

We have found from extensive testing that stages 1-3 are not critical to the sampler burning in; indeed the method can be started from stage 4 using a random prior sample and still converge to the same posterior distribution each time. Stage 1 primarily serves the purpose of aiding with outlier removal, which is a serious problem in differential time data. Stages 2-3 simply help the sampler to be a bit more stable and burn in faster. They ultimately cut down on compute time and also make it easier to get the hyperparameters in the right range.

Just like with all Bayesian inference problems \citep[e.g.]{duputel_accounting_2014}, having a credible probability model is essential for the results to be meaningful, and uncertainty estimates to reflect what we believe to be the truth. In SPIDER the prior distributions are critical for keeping the injected noise from pulling the solution in directions where the forward model has effectively no sensitivity. These null-like directions are thus dampened out making the solution well behaved. The Bayesian formulation makes these assumptions explicit and allows for the results to be interpreted in light of them. If unreasonable uncertainties are included then the posterior is not going to be meaningful. Thus it is important to think carefully about the right values to use, some of which can be estimated with hold-out observations. For our catalogs, we estimated the $\sigma, \kappa$ using residuals at the end of Stage 1, and separately estimated $\ell, p$ by fitting eq. \ref{eq:correlation_decay} to correlation diagrams like in Fig. \ref{fig:shared_error} for each event. The remaining inversion hyperparameters that may require tuning are the learning rate, batch size, and thresholds for the differential times; however, in this study, we have used nearly the same values for all four catalogs and thus we have some preliminary reason to believe that they may not need to be changed substantially.   

There are several metrics that are desirable to track to ensure the sampler is performing effectively. The first is the effective sample size, a standard metric in MCMC that is based on the integrated autocorrelation time; the effective sample size looks at the autocorrelation function for each parameter and determines how many samples can be considered effectively independent. The second metric is the ratio between the variance of the stochastic gradient to the variance of the Langevin injected noise. This ratio depends linearly on the learning rate theoretically and a good rule of thumb is to target a range of $0.1-1.0$. If it is larger than $1.0$, the sampler may never exit the optimization phase because the injected noise will be too weak or the posterior bias in Wasserstein distance may be large. If it is less than $0.1$, the sampler may move too slowly across the posterior. Finally it is important to verify that the loss and parameters have reached a stationary phase over many epochs, without any systematic drifting remaining.

The framework outlined in this paper is easily adaptable and can incorporate future improvements to SGLD from within machine learning, or alternative forward models. For example, there currently exist variants of SGLD including those based on Hamiltonian Monte Carlo \citep{chen_stochastic_2014}, adaptive drift terms like moment or Adam \citep{kim_stochastic_2022}. SPIDER itself only requires a differentiable forward model, thus does not strictly require an EikoNet; for example a PyTorch differentiable interpolation algorithm, neural field representation trained with supervision, or Neural Operator could all be swapped in to replace the forward model.

%
%

\section*{Open Research Section}
All data and code used for this paper are available from Zenodo \cite{ross_2025_16930266}.
\section*{Conflict of Interest Statement}
The authors have no conflicts of interest to disclose.
\section*{Acknowledgments}
The authors are grateful to Jannes Münchmeyer and an anonymous reviewer for their valuable reviews of the manuscript. This study was supported by the David and Lucile Packard Foundation and US National Science Foundation.

\bibliographystyle{unsrtnat}
\bibliography{references,refs2}

\end{document}